# Deformation Mechanisms in High Entropy Alloys: A Minireview of Short-Range Order Effects

Novin Rasooli[a]     Wei Chen[b,c,*]     Matthew Daly[a,*]

[a]Department of Civil, Materials, and Environmental Engineering, University of Illinois Chicago – 842 W. Taylor St., 2095 ERF (MC 246), Chicago, IL, 60607, United States

[b]Department of Mechanical, Materials, and Aerospace Engineering, Illinois Institute of Technology, Chicago, IL 60616, USA

[c]Department of Materials Design and Innovation, State University of New York at Buffalo, Buffalo, NY 14260, USA

**Abstract**

The complex atomic scale structure of high entropy alloys presents new opportunities to expand the deformation theories of mechanical metallurgy. In this regard, solute-defect interactions have emerged as critical piece in elucidating the operation of deformation mechanisms. While notable progress has been made in understanding solute-defect interactions for random solute arrangements, recent interest in high entropy alloys with short-range order adds a new layer of structural complexity for which a cohesive picture has yet to emerge. To this end, this minireview synthesizes the current understanding of short-range order effects on defect behavior through an examination of the key recent literature. This analysis centers on the nanoscale metallurgy of deformation mechanisms, with the order-induced changes to the relevant defect energy landscapes serving as a touchstone for discussion. The topics reviewed include dislocation-mediated strengthening, twinning and phase transformation-based mechanisms, and vacancy-mediated processes. This minireview concludes with remarks on current challenges and opportunities for future efforts.

---

[*]Corresponding authors: wchen226@buffalo.edu (W. Chen);  mattdaly@uic.edu (M. Daly)

---

**1. Introduction**

High entropy alloys (HEAs) are typically characterized by the presence of more than four elements in nearly equal atomic ratios and have risen as a fascinating class of materials with unique deformation behavior and defect properties.[1,2] At the atomic scale, these characteristics are underpinned by the signature structural feature of HEAs: a concentrated solid solution topology constructed from complex chemical arrangements patterned on high symmetry crystallographic lattices. Here, the traditional notion of solute and solvent is broken, and each atom can be viewed as a solute embedded in an effective medium that represents the surrounding environment.[3,4] This unconventional concept in alloy design has resulted in new and remarkable mechanical metallurgy, which includes unusual solid solution strengthening properties,[4] excellent high temperature strength,[5] and exceptional fracture toughness in cryogenic environments.[6,7] Beyond this collection of properties, the variable solute environment has redefined interpretations of classic mechanical metallurgy concepts such as the Peierls barrier,[4,8] the vacancy formation and migration energy,[9] and planar fault energies,[10–12] where overlaps in the length scales of defect structure and solute patterning introduces stochastic fluctuations to defect energies.

Although nearly 20 years have passed since the initial reports of HEAs,[13,14] these systems continue to attract great interest within the community, with more than 8000 peer-reviewed articles published since 2019 (see Figure 1). Much of the early literature in HEA research has focused on the analysis of solid solutions with random solute arrangements. This perspective largely stems from initial exploration of the CrMnFeCoNi system, where decomposition from a solid solution occurs only after extremely prolonged periods of annealing.[15] However, as the science of HEAs has matured and investigators examine an ever-increasing set of systems, an understanding that random (ideal) solid solutions are rare has begun to permeate the community. Indeed, recent activity in HEAs has witnessed a growing interest in investigations of short-range order (SRO) as a pathway to study the effects of non-random solute arrangement. As shown in Figure 1, interest in SRO is rising quickly, with the number of related articles increasing more than five-fold between 2019 and 2023 (25 to 135 articles, 2023 values annualized from September 2023). From a mechanical metallurgy perspective, the notion of SRO raises new questions regarding the impact of solute arrangement on the emergence of deformation mechanisms, with mixed reports of its importance emerging in the recent literature. For instance, some experimental studies have reported strong coupling of SRO to a wide range of mechanical properties including the tensile yield strength,[16] hardness,[17] and dynamic mechanical properties.[18] In each case, the strong effect of SRO on defect behavior is used as a common theme to rationalize observations. Conversely, other investigators have offered alternative explanations for trends in mechanical property data. For example, Yin et al.[19] showed that the yield strength for contemporary data on CrCoNi is well-predicted by random solute strengthening theory and contributions from SRO are likely negligible. Zhang et al.[20] further examined this literature discrepancy by separately studying the bulk and nanomechanical properties of CrCoNi. Here, the investigators report a limited effect of SRO on bulk strength but a significant influence on nanoindentation pop-in loads, which leads to the interpretation that SRO serves as a weak obstacle to aggregate dislocation glide but has a strong coupling to dislocation nucleation. Yet, as highlighted in these examples, a cohesive understanding of SRO effects has yet to emerge from the community.

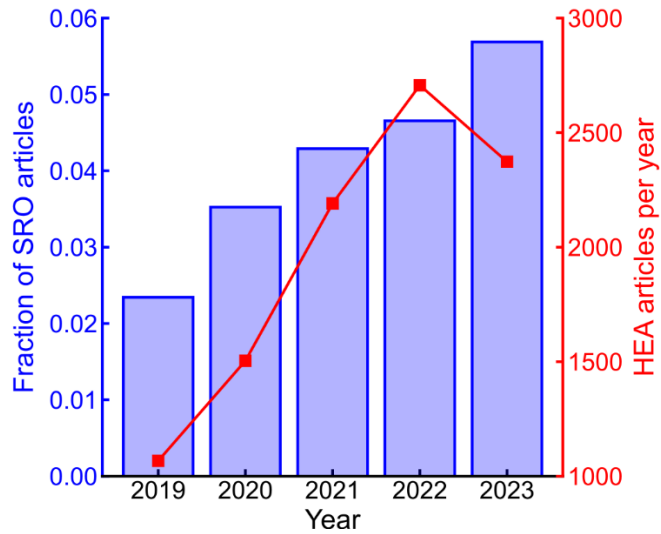

**Figure 1:** The increase in the fraction of peer-reviewed journal articles related to SRO effects in HEAs from 2019-2023 (in blue). The number of HEA articles published per year (in red). Data is obtained from the Web of Science with the values for 2023 annualized based on a search performed in September 2023.

Given the significant scientific interest and rapidly evolving understanding of SRO in HEAs, we have prepared this minireview to synthesize current understanding of its effects on deformation mechanisms. To supplement the relatively small (but quickly increasing) number of studies on this topic, we also discuss relevant analogous concepts available in the concentrated solid solution and medium entropy alloy (MEA, i.e., near-equimolar ternaries) literature. To position the effects of SRO on HEA deformation mechanisms within the fundamentals of defect metallurgy, the interplay of SRO with the relevant solute-defect interactions serves as a focal point for discussion. Here, the potential energy landscape (PEL) concept is leveraged as it provides a physical representation of these interactions at the atomic length scales relevant to solute patterning. Indeed, several studies have employed the PEL concept to investigate defect metallurgy phenomena such as solute embedding energies,[10] solute-dislocation interactions,[4,21] and vacancy formation / migration[22,23] kinetics in HEA systems. This article proceeds first with a description of methods to parameterize SRO and an overview of its impact on the structure of the PEL. This discussion is followed by an examination of SRO effects on deformation mechanisms in HEAs and their coupling to changes in the relevant PEL. This article closes with brief comments on the outlook for this topic. As SRO effects represent an emerging topic in the study of HEAs, the objective of this minireview is to synthesize the salient insights to help guide future research efforts with a scope that covers a modest list of key reports in recent literature.

## 2. Structural and defect metallurgy concepts

### 2.1. Parameterizing short-range order

Short-range order refers to the local deviations from a random arrangement of solute in a solid solution. SRO appears under several analogous terms in the literature, including local chemical ordering, chemical short-range order, and mechanically-derived short-range order, with the latter two terms signifying different processing routes for SRO formation (i.e., thermal or mechanical, respectively[24]). For the purposes of this report, we consider each term interchangeable in a structural sense, and for clarity will simply use the term SRO. Within the context of crystalline solids, SRO can be derived from correlations between atomic species at interatomic distances that are fixed by lattice symmetry, with random solid solutions (RSS)s serving as a benchmark. The thermodynamics of mixing favor RSSs when entropic effects dominate, whereas favorable enthalpic interactions drive the formation of specific solute arrangement patterns between species at lattice coordination sites leading to long-range order (LRO). SRO emerges between these extremes where neither entropic nor enthalpic effects dominate the thermodynamics of the solution. While the extent to which ordering effects can be considered short-range remains somewhat ambiguous, it is generally accepted that the distances over which solute interactions contribute significantly to system-level properties (e.g., total internal energy) serves as a useful guideline. Semi-quantitative methods to demarcate SRO from LRO have been described in the literature.[25] For instance, the system-level site replacement parameter of Bragg and Williams[26] serves as a classic example of LRO convergence criterion for distance-resolved pair correlation calculations.

Several parameters exist in the literature to quantify SRO, from which the Warren-Cowley (WC) parameter is best known.[25,27] In its original presentation, the WC parameter, $\alpha_{ij}^m$, is used to describe pair correlations in binary systems. It is formulated by normalizing the conditional probability of finding a solute of type *j* around an atom *i* at a distance of neighbor shell *m* against the probability of random arrangement:

$$\alpha_{ij}^m = 1 - \frac{p_{ij}^m}{c_j} \qquad (1)$$

where $p_{ij}^m$ is the conditional probability of the solute pairing in the $m^{th}$ neighbor shell, and $c_j$ is the overall composition of element *j* (i.e., the probability of random arrangement). Attraction and repulsion of solute are directed by the values of $\alpha$, with a random arrangement existing for $\alpha = 0$. An RSS can be established for systems where $\alpha^m = 0$ over a sufficient number of neighbor shells, with the distance thresholds on the order of the lattice parameter seeming reasonable. The interpretation of the extremes of the WC parameter is situational but offers important metallurgical insight into patterns in solute arrangement. For instance, when comparing dissimilar species (i.e., $i \neq j$), $\alpha_{ij} \to 1$ (m=1) indicates attraction of similar solutes (as $p_{ij} \to 0$). Conversely, attraction of dissimilar solutes emerges at the negative limit of the WC parameter, where $\alpha_{ij} \to \frac{c_j-1}{c_j}$ as $p_{ij} \to 1$. Figure 2 provides a schematic representation of SRO in a binary system. Despite its convenient form, the nature of its determination requires a homogenization of pair correlations over a sampling volume, which raises ambiguity in the interpretation of solute organization. That is, the WC parameter provides the statistical SRO in alloy systems, and not an explicit or unique description of chemical structure. As discussed by Owen et al.,[28] identical WC parameters can be calculated from large-box models of systems exhibiting microheterogeneities of distinct ordering embedded within a disordered matrix (see disperse and microdomain ordering in Figure 2). These microheterogeneities have been observed in binary systems[29,30] and can be presumed to present in HEA systems. Therefore, relationships between SRO-dependent physical properties and the WC parameter must be drawn with caution. This consideration becomes particularly relevant in studies where SRO is measured directly by large-box statistical approaches and then used as a collective parameter to benchmark changes in mechanical behavior.

The WC parameter appears in several equivalent forms in the literature, where different definitions of pair probabilities are selected depending on their convenience of calculation (e.g., joint or overall combinatorial probabilities of $i$ and $j$ pairs[11,31]). As relevant to HEAs, the original WC parameter has been expanded to multicomponent systems by De Fontaine.[32] Ceguerra and co-workers[33,34] further generalized the descriptor to analyze sets of solutes rather than pairs of atomic species. More recently, Goff et al.[35] used the cluster expansion formalism to develop a multipoint parameter that extends descriptions of SRO beyond pair correlations (e.g., to three- and four-point interactions). In contrast to the WC parameter, which is restricted to describing pairwise relationships, Goff's multipoint parameter describes the frequency of observation for a specific solute cluster with respect to an RSS. However, each of these advancements share common elements with the WC parameter from which they are developed. That is, each parameter uses random arrangement as a benchmark to quantify the degree of SRO and the original WC parameter can be recovered when analysis is restricted to pair correlations in binary systems. In addition to WC-based parameters, other SRO descriptors have been provided (e.g., Refs.[36,37]). However, like the WC parameter, these alternatives use the pair correlation of the random solid solution as a benchmark for local order.

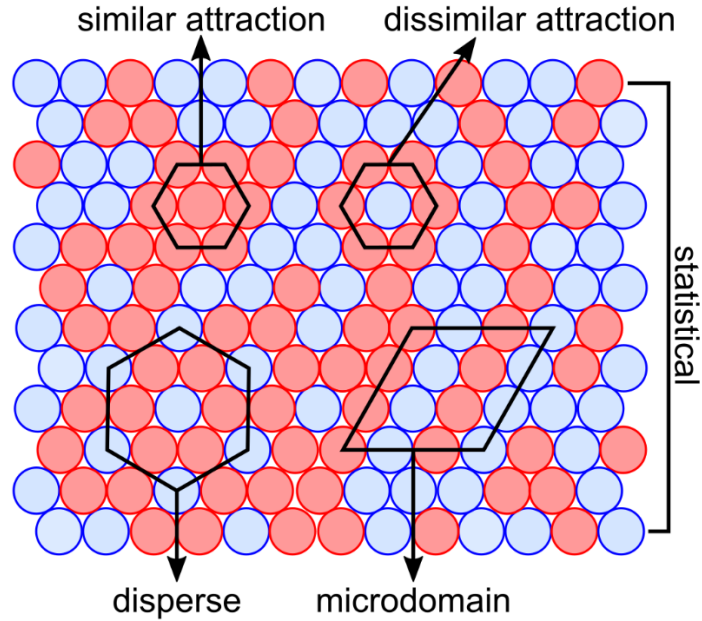

**Figure 2:** Structural representation of SRO in an equimolar binary alloy schematic. Examples of similar and dissimilar attraction are depicted for the nearest neighbors in the {111} plane of a face-centered cubic (FCC) crystal. Examples of different types of solute ordering in large-box models are shown using the nomenclature of Owen et al.[28] In statistical SRO, the solute mixing is homogeneous and individual neighbors are stochastically selected. This form of SRO is most closely represented by the WC parameter. Microheterogeneities may be embedded within the statistical SRO with ordering that diverges significantly from the statistics of the large-box. Owen et al.[28] refer to these regions as disperse and microdomain, with the local change in stoichiometry in the former arrangement being the distinguishing characteristic. Description of SRO using the WC parameter can mask the existence of these microheterogeneities.

Experimental methods to measure SRO parameters include scattering-based approaches using X-ray and neutron sources, absorption spectroscopy by the extended X-ray absorption fine structure (EXAFS) analysis, electron microscopy techniques, and atom probe tomography (APT) methods. Due to their long history in structure identification, scattering measurements represent the earliest pathway to measure SRO. Here, total scattering analysis has emerged as a method to directly quantify pair correlations between solute atoms in HEAs.[38] In contrast to Bragg scattering, which inherently considers only the long-range average structure of a material, total scattering examines both Bragg and diffuse contributions to collected measurements, which enables quantification of SRO. Total scattering can be analyzed using both X-rays and neutrons and has witnessed a resurgence in recent years due to the development of high resolution (i.e., high-Q) sources.[28,39] As shown in Zhang et al.,[40] EXAFS analysis can be used to complement total scattering measurements to detect SRO in systems where constituents exhibit similar scattering behaviors. In addition to scattering-based techniques, atomic-resolution imaging methods such as those available in transmission electron microscopy (TEM) have been applied to examine SRO in MEAs. For instance, diffuse superlattice reflections in electron diffraction patterns combined with energy-filtered dark field imaging revealed SRO in CrCoNi[17] and VCoNi[41] in recent high visibility studies. However, determination of SRO by superlattice reflections remains controversial, as some expected reflections are notably absent in reports and alternative sources for diffuse reflections exist, which include small, symmetry breaking lattice defects.[42] APT provides a three-dimensional, atomic scale measurement of solute arrangement. These datasets are particularly attractive as SRO parameters can be directly calculated from the reconstructed tomographic data. While attempts to measure SRO by APT often suffer from uncertainties due to poor detection efficiencies (~57%) and limits in lateral resolution, Monte Carlo-based techniques are available to partially reconstitute incomplete APT data.[33] New advancements in instrumentation promise considerable improvements to SRO parameter measurement with next generation APT systems offering detection efficiencies up to ~80%.[43]

### 2.2. Energy landscapes in high entropy alloys

The potential energy landscape (PEL) is a representation used to understand the effects on potential energy after a configurational change within a system. Within the context of defect metallurgy, these potential energy changes are often described over spatial coordinates and are quantified in terms of the excess energy introduced through the nucleation or motion of a defect.

The PEL concept is therefore instrumental in studying the fundamental processes that underpin deformation mechanisms. Perhaps the simplest example of the PEL is that of the cohesive energy (Figure 3a), which describes the energy required to remove an atom from its lattice site to an infinite separation. More complicated PELs exist, with those related to the motion of vacancies and interstitial atoms, and the glide of dislocations presenting as common examples. Figure 3 presents some examples of PELs for common defect processes. As shown in the figure, many PELs exhibit undulations whose local basins and peaks separate the stable and metastable configurations of a system. Indeed, these peaks represent the energy barriers resisting defect processes and their quantification is central to theoretical efforts in defect metallurgy. Several computational tools exist to measure energy barriers, which include direct atomistic simulations, reaction coordinate mapping (e.g., as in the nudged elastic band method[44,45]), or landscape sampling techniques.[46–48]

Within the context of HEAs, concentrated solutions add a layer of chemical complexity to the topography of the PEL. Whereas in pure metals the PEL exhibits regular undulations that correspond to the underlying symmetry of the crystal lattice, the variability in the solute environment creates a roughened, spatially heterogeneous topography (e.g., Figure 3b and c). This is markedly different from the behavior in dilute solutions, where roughening of the PEL certainly occurs, but arrives with a predictable shape within the backdrop of a pure solvent environment. This effect of the concentrated solute environment on the PEL is perhaps most easily understood through the cohesive energy landscape, where fluctuations arise only due to variations in the local chemical structure (Figure 3a). PELs related to lattice defects exhibit similar roughened topographies that emerge from variations in the solute-defect interaction energies. For instance, Curtin and co-workers[4,21,49] have developed a theory to predict the statistics of glide barriers arising from solute-dislocation interactions in randomly arranged HEAs. Related studies focusing on the statistics of defect-related PELs are emerging, including those focused on vacancy formation,[50–53] migration,[22,50–53] and the planar fault energy landscapes.[10,12,54]

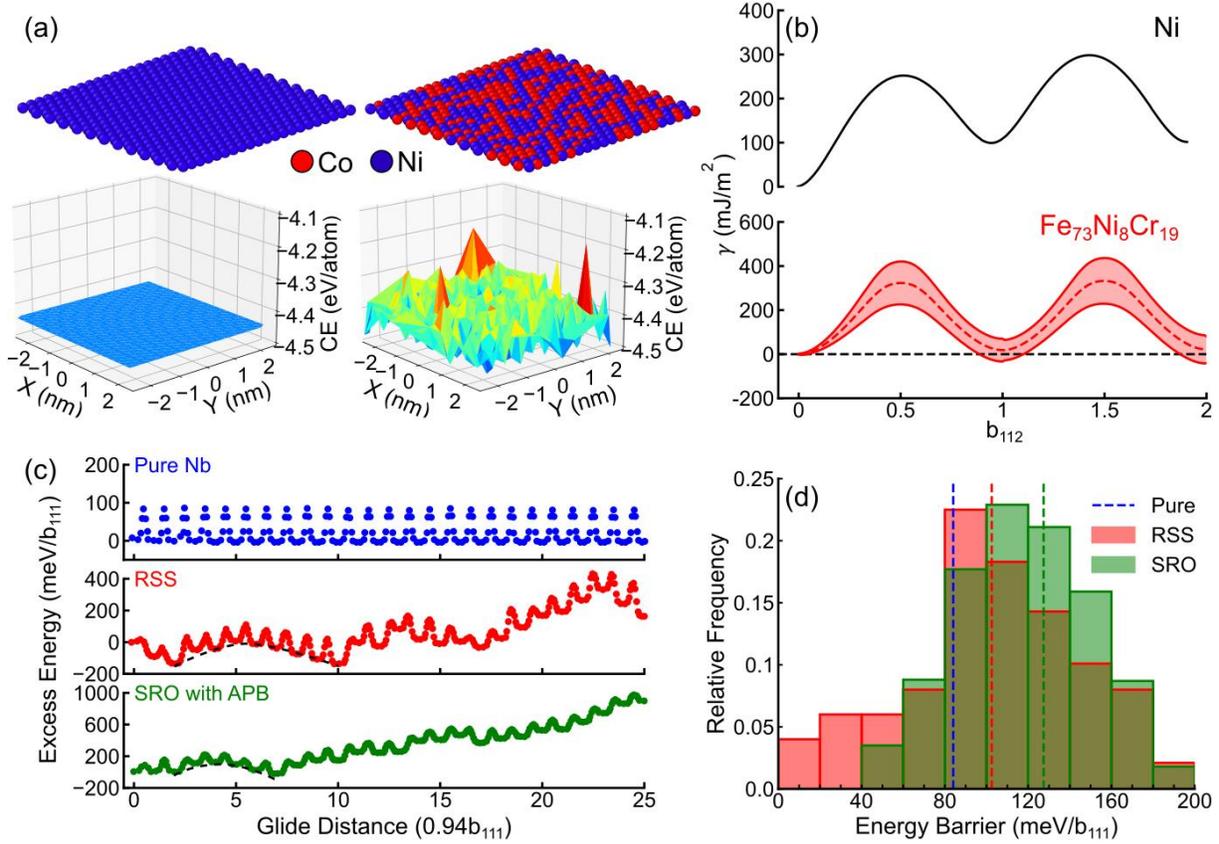

**Figure 3:** Examples of PELs for cohesive energy (a), generalized planar fault energies (b), and dislocation glide barriers (c). (a) The local variations in cohesive energy (CE) are compared in the {111} plane of a FCC equimolar CoNi RSS against a pure nickel reference.[10] (b) The planar fault energies ($\gamma$) are plotted against the lattice shear in units of the <112> FCC Burger's vector ($b_{112}$) for pure Ni and a $Fe_{73}Ni_8Cr_{19}$ RSS (by mole).[10,55] The statistical fluctuations in the FeNiCr system are shown, with the dashed line representing the average planar fault energies and the upper and lower bounds enclosing one standard deviation. (c) The Peierls barriers for screw dislocations in pure Nb, and a MoNbTaW HEA under RSS and SRO arrangements. The glide distance is plotted in units of the <111> BCC Burger's vector ($b_{111}$). The Peierls barriers for the pure material are constant, as expected, and exhibit variations for the RSS and SRO conditions. The pathway connecting two metabasin valleys in the RSS and SRO data is traced in dashed stroke. The baseline of the SRO excess energy climbs steadily due to the growth of an antiphase boundary with each successive slip step. A histogram showing the distributions of the RSS and SRO Peierls barriers is plotted in (d) with the average barrier marked by the relevant dashed line.[56,57]

Most literature related to changes in the PEL due to the concentrated solute effect are restricted to an assumption of a randomly arranged solid solution. In addition to the added structural complexity, SRO also introduces antiphase boundary (APB) energies into the PEL when lattice shearing underpins the landscape construction (e.g., as in dislocation glide). As shown in Figure 3c, the effect of the APB appears as a tilt in the PEL, which arises from the unfavorable energetics of APB growth with progressive dislocation glide. The PEL for SRO arrangements share some common features with RSS configurations, such as the existence of metabasins, which identify the characteristic length scale of waviness in dislocation lines. These larger scale hierarchical PEL features form the theoretical basis of the metabasin-hopping strengthening model proposed by Curtin and co-workers.[4,21,49] For the studies that have reported on the effects of SRO on the PEL, an understanding of the effect on defect energies is still emerging. For instance, Wang et al.[56] show a narrowed distribution of Peierls barriers for screw dislocations in body-centered cubic (BCC) HEAs with SRO when compared to the barrier statistics for a random reference alloy (Figure 3c and d). A similar result is reported in the same system by Yin et al.,[58] who found an inverse relationship between the degree of SRO and statistical variance in the core energies of screw dislocation dipoles. These observations align with the understanding of the intermediary effect of SRO on structure. That is, variability in the PEL is maximized and minimized at the extremes of random and ordered arrangements, and SRO presents as a gradient in variability. However, studies from Cao and co-workers[22] report a broadening of the distribution of vacancy migration barriers in two MEA systems due to SRO. This increase in the variability of PEL barriers is also reported for vacancy migration in a study from Zhao.[53] With respect to the narrowing observed in dislocation glide barriers, one possible explanation for these differences is due to the directionality of the deformation mechanisms. That is, dislocation glide processes are generally studied in forward steps only, which naturally follows from operation of the deformation mechanism under an applied shear stress. Conversely, vacancy migration is bidirectional (i.e., forward and backward). While a forward migration barrier may increase due to PEL roughening from the preferred pair correlations, a backward migration barrier can be similarly reduced due to energy penalties associated with breaking SRO. On aggregate, the dispersion of energy barriers between forward and backward vacancy migration steps is wider than in a RSS, which leads to a broadening (and separation) of the distribution. Nonetheless, as this literature is still emerging, it will likely take some time for a complete understanding of SRO effects on the PEL to emerge from the community.

## 3. Dislocation-mediated strengthening in HEAs with SRO

Strengthening in HEAs results mainly from the interactions between dislocations and fluctuations in the local solute environment.[4] Compared to a RSS, SRO alters the distribution of elements in a HEA, creating stronger chemical bonds and more compact local configurations near dislocations. While solid solution strengthening due to the size mismatch of constituent elements is often dominant in determining the strength of a HEA, the local chemical and structural changes from SRO can introduce additional contributions. In these alloys, SRO usually reduces the local hydrostatic strain from atomic size misfit and lowers the energy of the system. The complex interaction between the atomic misfit and SRO is therefore critical to understanding the effects of SRO on strength. In FCC HEAs, these two factors have been found to have an opposite effect on the critical resolved shear stress to unpin an edge dislocation. For screw dislocations, the effect of the atomic misfit is negligible and the critical stress increases with SRO.[59] However, it is often difficult to distinguish the effects of these two strengthening mechanisms on the mechanical properties of HEA. For example, Yin et al.[19] concluded SRO in CrCoNi has no systematic measurable effect on strength and that the high strength of CrCoNi can be understood solely from the misfit volumes and the elasticity of the random alloy.

The disruption of favorable SRO by the glide of a leading dislocation forms a 'diffuse' APB with an altered SRO. This boundary is correlated with a "glide softening" effect where trailing dislocations feel reduced resistance to glide. The reduced resistance to planar slip from glide softening can lower cross-slip and lead to delayed dynamic recovery or an increased twinning formation. A recent study in a concentrated FCC binary solution confirms the increase in the cross-slip energy barrier that is dependent on both the degree of SRO and the existence of a diffuse APB.[60] Meanwhile, although SRO often reduces the overall cross-slip rate, the existence of a diffuse APB also suggests the high probability of repeated cross-slip at the same site. This tendency can thus create correlated cross-slips, double cross-slips, and eventually dislocation pile ups that further increase the work hardening of the alloy. The effect of SRO on cross-slip as a rate-limiting process is likely a critical factor in tuning the deformation mechanisms of HEAs and warrants further investigation.

Most BCC metals and their dilute alloys exhibit obvious ductile-to-brittle transitions at intermediate temperatures. In these conventional BCC alloys, non-screw dislocations can glide easily, with the more sluggish screw dislocations only contributing to plasticity appreciably at higher temperatures, causing brittleness at intermediate and low temperatures. However, the strengths of some BCC HEAs have shown a surprisingly weak temperature dependence, with only a gradual decrease in strength at higher temperatures.[61] In a random BCC MoNbTi HEA, while edge dislocations still account for most slip activity, the barriers for kink-pair nucleation on screw dislocations are also lowered because of the varying dislocation core structures in the alloy.[62] The kinks on screw dislocations with an edge character can glide on various planes, even on non-{110} slip planes with lower packing densities, improving ductility at intermediate temperatures.

While SRO has been reported in some studies to have a significant impact on the strength of HEAs, correlation of strengthening mechanisms with dislocation evolution remains controversial. Some progress has been made in revealing the fundamental defect metallurgy, with SRO-induced effects on dislocation energy barriers, dislocation structure, and glide kinetics being reported. For instance, Beyerlein and co-workers[63] examined the homogeneous nucleation of Shockley partial dislocations in a CrCoNi MEA by molecular dynamics simulation. Here, an increase in the homogenous nucleation stress and slower dislocation glide kinetics are reported for SRO systems when compared to an RSS benchmark. This observation aligns with PEL data from related computational studies and finds partial agreement with the available experimental data. For instance, Cao and co-workers[56] report an average increase in the kink-pair nucleation and Peierls glide barriers (Figure 3c and d) for screw dislocations in a MoNbTaW HEA. These increases are attributed to the energy penalty produced from the creation and growth of an APB during dislocation nucleation and glide. Yin et al.[58] complement this finding by showing a monotonic link between the degree of SRO and the APB energy in their ab initio study of this same system. The APB-related strengthening mechanism also serves as an interpretation for the increase in dislocation nucleation stresses in a CrCoNi MEA with SRO, as measured by nanoindentation pop-in experiments from Zhang et al.[20]

While there is an emerging consensus that SRO can increase dislocation nucleation stresses, property trends in bulk strength are less clear. For instance, although Zhang et al.[20] reveal SRO-induced increases in dislocation nucleation strengths by nanomechanical testing, they fail to observe any correlations in bulk yielding behavior. One possible explanation for this discrepancy is the differential influence of SRO on the varied rate-controlling mechanisms of dislocation-mediated plasticity. Within this context, Cao and co-workers[56] observed a suppression of kink-pair mechanisms in screw dislocations for the MoNbTaW SRO-HEA compared with a RSS benchmark, which arrives with a shift in preference towards kink-glide processes. Ritchie and co-workers[64] provide further analysis of this HEA system, with a concurring a report of reduced kink-pair kinetics in screw dislocations, but also an increased mobility for edge dislocations. Conversely, a recent computational study on a BCC MoTaTiWZr HEA reveals disproportionate increases in energy barriers for both edge and screw dislocation motion relative to a RSS benchmark, making the atomic scale mobility of edge dislocations similar to or even lower than screw dislocations (see Figure 4).[65] One complication in reconciling the existing literature is a lack of consensus on the fundamental processes by which dislocation glide proceeds, with traditional systems offering little useful guidance. For instance, due to the irregular line-structure of BCC screw dislocations, it remains unclear if glide is limited by kink-pair nucleation, kink-glide, Peierls, or metabasin-hopping mechanisms. A physical understanding of SRO effects on the bulk deformation mechanisms of HEAs therefore requires additional mesoscale investigations to reveal the aggregate impact of rate-limiting dislocation processes on defect structure evolution and strengthening.

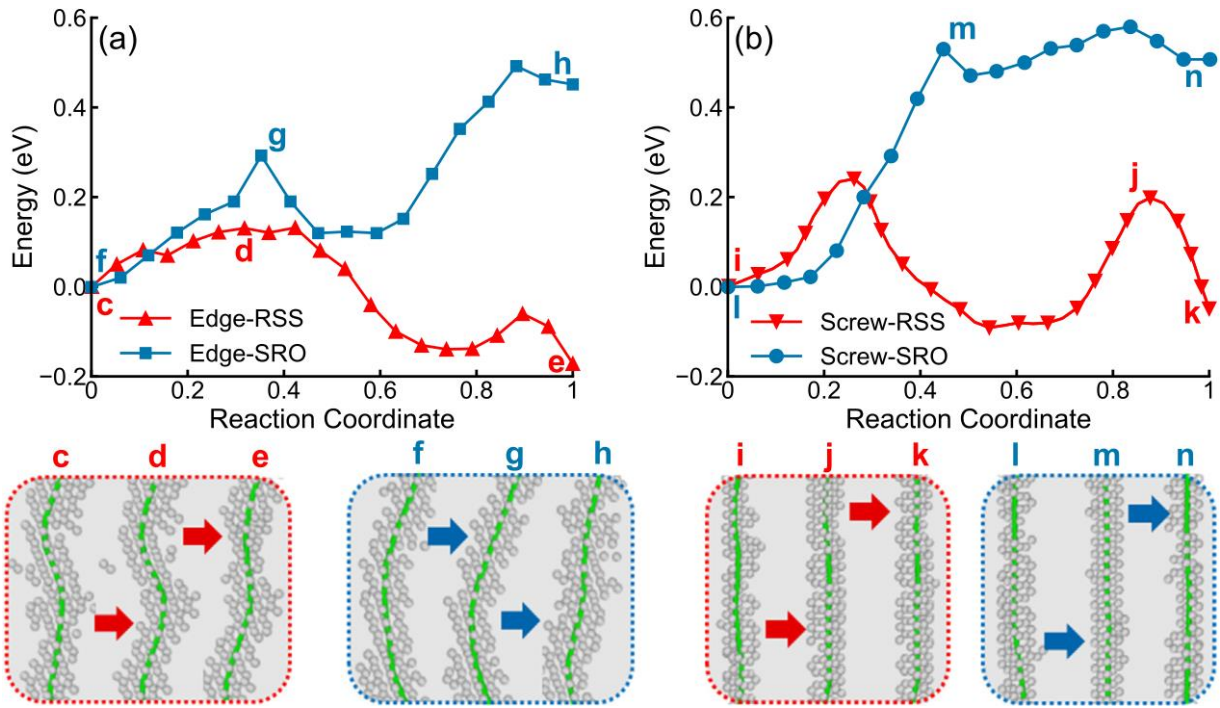

**Figure 4:** The effect of SRO on the minimum energy pathway for the core of an edge (a) and screw (b) dislocation motion in the MoTaTiWZr HEA compared to an RSS reference. SRO increases the excess energies for both edge and screw dislocation glide, but the effect is stronger for the edge dislocation, potentially making it less mobile than the screw dislocation.[65]

## 4. Twinning and phase transformation-based deformation mechanisms

Traditional strengthening mechanisms such as precipitation and solid solution hardening present with tradeoffs in ductility. By comparison, twinning and phase transformation-based deformation mechanisms accommodate plasticity through crystallographic changes that have been demonstrated as an effective strategy to overcome conventional strength-ductility trade-offs. In HEAs, metastability offers a pathway to tailor this transformation- or twinning-induced plasticity (TRIP/TWIP)[66,67]. During the deformation of TWIP/TRIP HEAs, martensite or twin formation provide alternative pathways for partial dislocation glide, while the phase or twin boundary reduces the dislocation mean free path, leading to the dynamic Hall-Petch effect. The competing deformation mechanisms in FCC HEAs enabled by metastability, such as dislocation glide, twinning, and martensitic transformation, can be predicted by the generalized stacking fault energy (GSFE) landscape and its variants (e.g., the generalized planar fault energy landscape, see Figure 3b). The intrinsic stacking fault energy on a GSFE curve, for example, is the excess energy associated with the formation of an intrinsic stacking fault by the dissociation of a lattice dislocation into two partial dislocations. The existence of SRO creates an additional barrier for the initial slip of an atomic layer. Nonetheless, once the slip takes place, it leads to the collapse of the local SRO, affecting both the further slip of the present layer and its neighboring layers (Figure 5a). To understand the evolution of stacking faults in HEAs with SRO, an extension of the GSFE concept was proposed to describe the effect of slip history and coupled multi-layer slip.[68] In the CrCoNi HEA, rather than a homogeneous distribution of slip in the atomic planes, SRO creates a more severe slip activity in certain atomic layers owing to the history-dependency of the intrinsic stacking fault energy. Furthermore, the history-dependent effect suppresses twinning and phase transformations for the CrCoNi HEA annealed at lower temperatures but promotes the formation of twins when the shear strain is increased.

In addition to impacting competition between mechanisms, the development of SRO in HEAs can facilitate local phase transformation under deformation and produce structural homogeneity similar to composite materials. In the CoCuFeNiPd HEA, a SRO-induced pseudo-composite microstructure has been reported to simultaneously enhance the ultimate strength and ductility (Figure 5b).[69] The microstructure consists of FCC-preferred clusters, BCC-preferred clusters, and the remaining indifferent clusters. Prior to the ultimate tensile stress, the HEA undergoes a localized FCC to BCC phase transition (Figure 5c). SRO in the HEA creates a local lattice distortion-induced instability where the phase transition can proceed without the necessity of nucleation of the BCC phase. The post-ultimate stress deformation of the HEA switches to partial dislocation slip and stacking fault formation within the FCC phase (Figure 5d). Overall, the indifferent clusters can be treated as the matrix, while the FCC-preferred clusters serve as the hard filler to improve the strength and the BCC-preferred clusters as the soft filler to increase the ductility. The surprising microstructure and mechanical properties of the HEA highlight the importance of SRO and offer new strategies to design high-performance structural alloys.

While an understanding of the impact of SRO on twinning and phase transformation-based deformation mechanisms is beginning to emerge, much of the fundamental defect metallurgy remains unclear. Indeed, the community has continued to grapple with difficulties in measurement and concerns regarding the applicability of fundamental defect metallurgy parameters to the examination of these deformation mechanisms. For instance, significantly different values of the stacking fault energy have been reported for several MEA/HEA systems, with values in the range of -62 to 22 mJ/m$^2$ in CrCoNi serving as one high profile example.[36,70–73] These differences are partially explained by the variations in solute arrangement in HEAs, which create a distribution of possible stacking fault energies that vary locally within a fault plane.[54,74] This notion of a 'local' stacking fault energy has important implications on deformation mechanism competition and the stability of the FCC phase. However, a clear understanding of this 'local' effect is complicated by additional difficulties in obtaining experimental measurements of stacking fault energies, which typically rely on measurement of the equilibrium splitting distances between partial dislocations. As demonstrated by Ghazisaeidi and co-workers[75], the traditional force balance approach used in splitting distance calculations overlooks a signature feature of HEAs. That is, the strong dislocation-solute interactions that are key to HEA solid solution strengthening[4] also create significant drag on partial dislocations that are not captured by the force balance model. Furthermore, the length scale sensitivity of the statistical scatter in stacking fault energy values has been demonstrated in separate studies from Zhao et al.[74] and Daly and co-workers.[12] Here, the investigators have shown how the distributions degenerate to singular values when fault energies are sampled over areas measuring more than a few square nanometers. Models to define the relationship between length scale and variance in fault energies have also been proposed for HEAs under RSS and SRO arrangements.[10,11] Within the context of twinning and phase-transformation based mechanisms, these findings are significant as they define the critical length scale over which fluctuations in fault energies influence defect processes. Phrased differently, fluctuations in energies are anticipated to be consequential to deformation mechanisms when there is a significant overlap with the length scale of the relevant defect structure (e.g., as in twin growth where a partial dislocation glides on faulted slip planes).

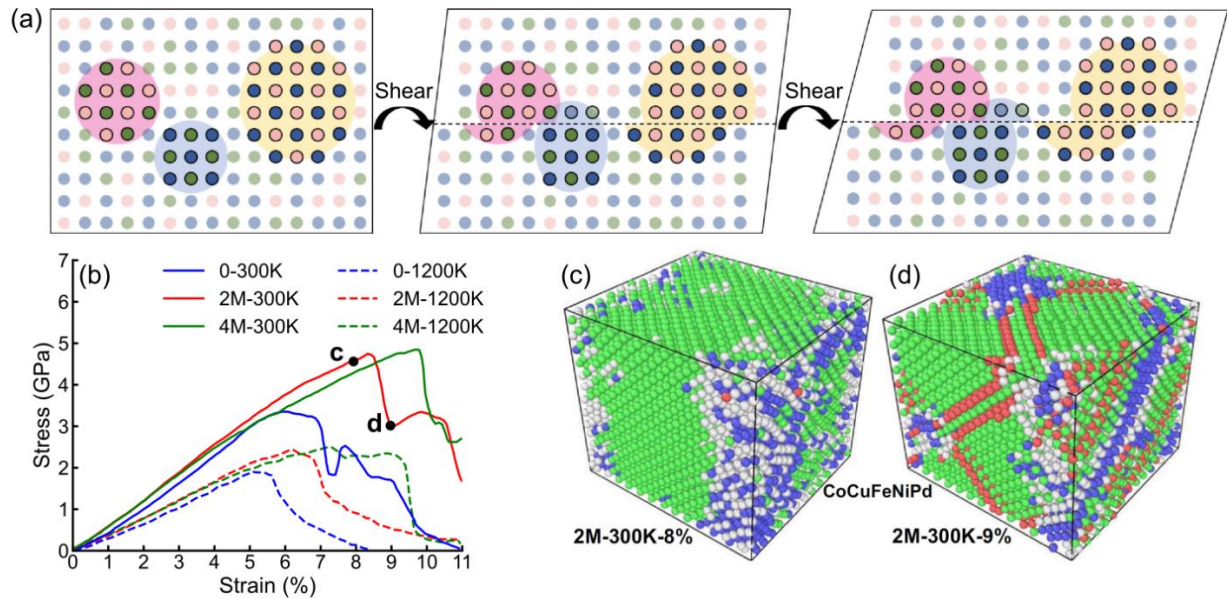

**Figure 5:** (a) Collapse of SRO during the creation of a stacking fault in a HEA, resulting in a added excess energy in the system.[68] (b) Molecular dynamics simulated stress-strain curves for CoCuFeNiPd HEA with 0, 2 million, and 4 million (M) equilibration steps for varied degree of SRO at different testing temperatures. The samples with stronger SRO show higher ultimate stress. (c) and (d) Atomic configurations for the 2M sample at 8% and 9% strain. Green, blue, and red atoms indicate FCC, BCC, and hexagonal close-packed (HCP) environments. The formation of stacking faults (HCP layers) releases local stress, increasing the ductility of the HEAs.[69]

## 5. Vacancy-mediated processes: Irradiation damage and creep

As HEAs represent a promising class of materials for applications in extreme environments, their response to irradiation damage[76] and creep[77] is of great interest to the community. At the nanoscale, these mechanisms are driven by the aggregate operation of point defects, among which vacancies and self-interstitials are perhaps the most important. Revealing the fundamental defect metallurgy that underpins the formation, migration, and interaction of these point defects in the complex solute environment of HEAs is critical to understanding their impact on deformation processes. For the purposes of this section, we primarily focus on the vacancy defect due to its importance in both deformation mechanisms, but much of the discussion is also applicable to interstitials. In irradiation damage, the bombardment of a material with energetic particles causes displacement damage in the lattice, which creates vacancies among other defects. Within the context of the PEL, here the vacancy formation energy (VFE) is arguably the most important parameter as it quantifies the incipient damage caused by bombardment. Conversely in creep, deformation is defined by time-dependent flow of a material, which is significantly influenced by vacancy transport. Examples of vacancy-driven creep processes include dislocation climb (as in dislocation creep), solute drag creep, and boundary-mediated processes as in Nabarro-Herring and Coble creep. In each case, the vacancy migration energy (VME) is an important fundamental defect metallurgy parameter that underpins the kinetics of the deformation process. Naturally, both the VFE and VME are sensitive to their solute environments. Therefore, the SRO-induced changes in the characteristics of the VFE and VME within a HEA solute environment serve as the focus for the discussion in this section.

Within the context of defect metallurgy, both the VFE and VME may be conceptualized within the framework of the PEL under an excess energy description, where the values of each defect energy arrive from a comparison of the system potential energy at critical stages of the defect process. For the VFE, the excess energy calculation proceeds from the difference of cohesive energies before and after the creation of the vacancy defect. With the VME, its value is derived from the maximum in excess energy during a vacancy-atom swap. In each case, the excess energy arises due to the interaction of the distortion field around the defect with the crystal lattice. In pure materials and dilute alloys, the VFE and VME are viewed as constant due to the regularity of the solute environment. In HEAs, however, the variable solute environment causes fluctuations in these defect energies, leading to distributions in values of VFEs and VMEs. The statistics of these VFE and VME distributions have been reported in several studies.[51,78–85] These variations in vacancy defect energies have been interpreted experimentally through the changes in activation volumes measured during nanoindentation experiments, as discussed in Nieh and co-workers.[86,87]

The effects of SRO on the vacancy defect energies may be understood from a similar gradient-based perspective as taken for structural considerations. That is, systems with SRO exhibit a mixing of LRO and RSS behaviors, with the distribution of vacancy defect energies narrowing as the degree of SRO approaches the LRO limit. Some evidence of this behavior is available in the works of Zhao[53] and Xing et al.,[22] which compare the distributions of VMEs for various HEAs under RSS and SRO arrangements. Here, both works report a broadening in the distribution of VMEs in the SRO configuration. As noted by Xing et al.,[22] this broadening emerges from a bias towards lower reverse barriers in vacancy migration that arise due to the unfavorable energetics of SRO-breaking vacancy transport (see Figure 6a-c). Therefore, in the limit where SRO becomes LRO, the distributions of forward and reverse migrations degenerate to VMEs associated with a specific vacancy-atom swap in an ordered solute environment.

While a complete understanding of the effects of SRO on vacancy properties is still emerging, new studies have emphasized the sensitivity of vacancies to SRO as a pathway for tunable behaviors. For instance, Osetsky et al.[79] have overviewed the sensitivity of point defects energies to the local chemical environment, which can be accessed through control of SRO. Diffusivity (as a companion property to vacancy migration) represents one such property of great interest due to its relevance in irradiation resistance and creep. New studies have revealed the beneficial role that SRO holds in slowing diffusion kinetics through defect trapping,[22,53,80,88,89] which can be explained by SRO-induced increases to forward vacancy migration barriers (see Figure 6c). However, Manzoor and Zhang[80] note that the presence of SRO does not ensure lower diffusivities. That is, while SRO may diminish vacancy migration, total diffusion may rise in aggregate due to SRO-induced increases to the equilibrium vacancy concentration. Furthermore, some studies have shown how vacancy migration pathways can become restricted to local regions of a lattice due to SRO roughening of the PEL.[22] Therefore, the rate of diffusion becomes less significant to material flow as vacancy defects become trapped over a length scale defined by the SRO. This interpretation aligns with a molecular dynamics study of creep resistance in the CrCoNi system from Huang et al.[90] Here, the investigators report a significantly higher activation energy under a power law creep model when comparing SRO samples to an RSS benchmark (see Figure 6d). Further examination of the solute-specific diffusivities under SRO reveals interesting behavior (see Figure 6e). Namely, the activation energy of power law creep was found to be close to the activation energy of Cr diffusion (1.73 and 1.71 eV, respectively), which is significantly higher than diffusion energies of Ni and Co (1.21 and 1.55 eV, respectively). Although diffusion of Ni and Co solute exhibits faster kinetics, creep is rate-limited by Cr migration, which indicates a solute-specific local trapping of Ni and Co under SRO. Similar behaviors were not observed in the RSS reference, where solutes exhibit similar activation energies that match with power law creep values (0.62 – 0.69 eV). These findings highlight opportunities to leverage SRO as a pathway to tune the creep resistance of HEA systems.

In addition to enhancing creep performance, SRO presents an opportunity to improve the radiation resistance of HEAs. As discussed in Su et al., [91] elements with diverse chemical affinities, such as interstitial alloying solutes, can contribute to this strategy. SRO near interstitial atoms results in composition variations, lattice strain, and limited interstitial diffusion channels, which reduces void swelling after irradiation. From a fundamental defect metallurgy perspective, this roughens the PEL, which impedes the motion of fast-moving self-interstitials and their clusters. This hinders the migration of radiation-induced defects, restricting their movement to local areas and void creation is delayed by easier recombination of interstitials and vacancies.[91] This recombination-based mechanism for improved irradiation resistance is also reported in an experimental irradiation study of the CrCoNi MEA from Zhang et al.[92] Here, the investigators measure significantly lower defect densities (but similar defect sizes) over a wide range of irradiation dosing for samples with SRO relative to an RSS reference (see Figure 6f and g).

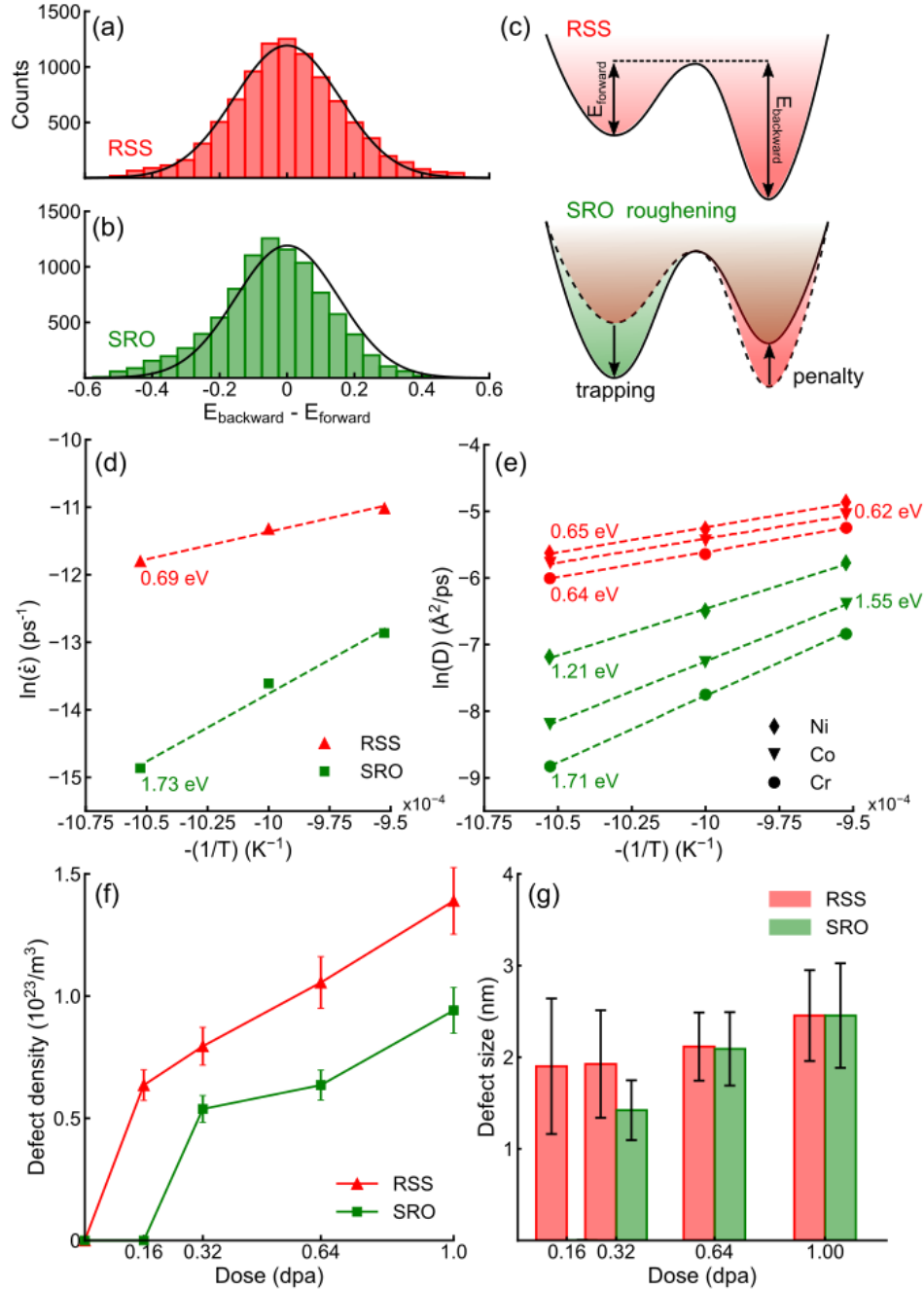

**Figure 6:** Difference between the backward and forward energy barriers to vacancy migration in a NbMoTa MEA under RSS (a) and SRO (b) arrangements. The gaussian distribution corresponding to the RSS data is plotted over both histograms to highlight the skew in the SRO condition. (c) SRO-induced roughening of the PEL for vacancy migration. Here, the energetics of SRO-breaking migration simultaneously increases the forward and reduces the backward barrier. (d) Molecular dynamics measurements of creep strain rate ($\dot{\varepsilon}$) in a CrCoNi MEA (RSS and SRO configurations) under a 0.5 GPa tensile load. A power law fit to the creep data is shown in dashed stroke, with the activation energies indicated. (e) Solute-specific diffusivities (D) in the CrCoNi MEA. The RSS and SRO data are presented in red and green, respectively. The activation energies for each process are also provided. The defect density (f) and defect size (g) of an irradiated CrCoNi MEA in RSS and SRO configurations. Data for each subfigure has been obtained from Xing et al.,[22] (a)-(c); Huang et al.,[90] (d) and (e); and Zhang et al.,[92] (f) and (g).

## 6. Outlook

The effect of SRO on deformation mechanisms in HEAs presents opportunities for new theory, design, and materials processing contributions to mechanical metallurgy. Yet, much remains unclear about the fundamentals of the defect processes that underpin deformation mechanisms. In this regard, significant challenges persist that we anticipate will guide efforts in the community. For instance, a cohesive understanding of the sequencing of defect processes that connects dislocation-mediated plasticity at the nanoscale with aggregate behavior in the bulk has yet to emerge. This knowledge gap is complicated by unclear trends in the impact of SRO on the energy barriers of fundamental defect processes, which challenges predictions of rate-limiting defect mobilities. While some progress has been made in relating SRO to the topography of the relevant PEL, much work remains to understand the varied influence of defect character and the strength of SRO coupling to the mechanical metallurgy. Defect modelling in HEA systems also presents scaling challenges. In contrast to random systems, where solute interactions can be averaged to model defect mechanics across length scales, ordered systems undergo changes in SRO parameters during deformation, which significantly complicates homogenization efforts. Obtaining computationally efficient methods to incorporate deformation history in the scaling of plasticity laws from the nanoscale presents one pathway to resolve this important issue.

**Author contributions**

**Novin Rasooli:** Data Curation; Visualization; Writing – Original Draft; Writing – Review & Editing. **Wei Chen:** Visualization; Writing – Original Draft; Writing – Review & Editing. **Matthew Daly:** Conceptualization; Data Curation; Funding Acquisition; Project Administration; Supervision; Visualization; Writing – Original Draft; Writing – Review & Editing

**Acknowledgements**

This material is based upon work supported by the National Science Foundation under Grant No. DMR-2144451 and DMR-1945380.

**Conflicts of interest**

There are no conflicts to declare.